\documentclass[12pt]{article}

\begin{document}

\title{\bf Addendum: Symmetries of the energy-momentum tensor}

\author{M. Sharif \thanks{e-mail: hasharif@yahoo.com}
\\ Department of Mathematics, University of the Punjab,\\ Quaid-e-Azam
Campus Lahore-54590, PAKISTAN.}

\date{}

\maketitle

\begin{abstract}
In recent papers [1-3], we have discussed matter symmetries of
non-static spherically symmetric spacetimes, static plane
symmetric spacetimes and cylindrically symmetric static
spacetimes. These have been classified for both cases when the
energy-momentum tensor is non-degenerate and also when it is
degenerate. Here we add up some consequences and the missing
references about the Ricci tensor.
\end{abstract}


\date{}

Recently, we have presented a detailed analysis of matter
collineations (MCs) for non-static spherically symmetric spacetmes
[1], static plane symmetric spacetimes [2] and cylindrically
symmetric static spacetimes [3]. We have discussed in detail the
matter symmetries for each of the metrics and have found the
corresponding constraint equations. In general, it is not easy to
solve these constraint equations even sometimes the solution of
the constraint equations may not exist. We have constructed some
examples which help us in exploring the difference between RCs and
MCs.

It is usually believed that matter and Ricci symmetries are the
similar symmetries and one can find MCs directly from the RCs.
However, this is not true in general. This has been shown in many
papers on this topic [1-9]. In this short communication, we
express this difference with examples. Further, we add up some
references missing in the papers [1-3] which should have been
inserted there.

Let $(M,g)$ be a spacetime manifold with signature $(+,-,-,-)$. It
is assumed that the manifold $M$, and the metric $g$, are smooth.
Einstein's field equations (EFEs), which relate the geometry and
matter, are given by
\begin{equation}
R_{ab}-\frac{1}{2}Rg_{ab}\equiv G_{ab}=\kappa T_{ab}, \quad
(a,b=0,1,2,3),
\end{equation}
where $\kappa$ is the gravitational constant, $G_{ab}$ is the
Einstein tensor, $R_{ab}$ is the Ricci and $T_{ab}$ is the matter
(energy-momentum) tensor. Also, $R = g^{ab} R_{ab}$ is the Ricci
scalar. It is obvious from EFEs that for vacuum spacetimes,
$R_{ab}=T_{ab}$ and consequently, RCs and MCs are similar in this
special case.

We define a differentiable vector field $\xi$ on $M$ to be a {\it
matter collineation} if $\pounds_\xi T_{ab}=0$ which can be
written in component form as
\begin{equation}
T_{ab,c} \xi^c + T_{ac} \xi^c_{,b} + T_{cb} \xi^c_{,a} = 0,
\end{equation}
where $\pounds$ is the Lie derivative operator, $\xi^a$ is the
symmetry or collineation vector. Since the Einstein tensor is
related to the matter content of the spacetime by the EFEs, the
investigation of MCs seems to be more relevant from the viewpoint
of physics. Here we would not give details of the calculations as
the procedure has been explicitly given in the papers [1-3].
Rather we would explore the difference of RCs and MCs for
non-static spherically symmetric, static plane symmetric and
cylindrically symmetric static spacetimes with the help of
examples.

The most general form of the metric for a spherically symmetric
spacetime is given by
\begin{equation}
ds^2=e^{\nu(t,r)}dt^2-e^{\mu(t,r)}dr^2-e^{\lambda(t,r)}d\Omega^2,
\end{equation}
where $d\Omega^2=d\theta^2+\sin^2\theta d\phi^2$. The surviving
components of the energy-momentum tensor are
$T_{00},~T_{01}~T_{11},~T_{22},~T_{33}$, where
$T_{33}=\sin^2\theta T_{22}$. We have found [1] that, for the
non-degenerate case, there exist either four, six, seven or ten
independent MCs in which four are isometries and the rest are the
proper. For the degenerate case, most of the cases give the
infinite dimensional MCs. The worth noting cases are those where
the energy-momentum tensor is degenerate but the group of MCs is
finite-dimensional, i.e., four or ten. Similar analysis has been
given in the paper [10] for the Ricci tensor. It can be seen from
the comparison of the two papers [1,10] that MCs and RCs turn out
to be the same but the constraint equations are entirely
different. For example, in the Einstein/anti-Einstein metric, we
obtain seven MCs [5] but RCs are infinite dimensional.

The metric for static plane symmetric spacetimes is given in the
form [11]
\begin{equation}
ds^2=e^{\nu(x)}dt^2-dx^2-e^{\mu(x)}(dy^2+dz^2).
\end{equation}
The surviving components of the energy-momentum tensor are
$T_{0},~T_{1},~T_{2},~T_{3}$, where $T_{3}=T_{2}$. When we solve
MC equations for the static plane symmetric spacetimes it turns
out [2] that the non-degenerate case yields either four, five,
six, seven or ten independent MCs in which four are isometries and
the rest are proper. We have also obtained three interesting cases
where the energy-momentum tensor is degenerate but the group of
MCs is finite-dimensional which are either four, six or ten.

Again when we compare the analysis given in the two papers [2,12],
it is concluded that RCs and MCs are similar but with different
constraint equations. We can construct some examples by solving
these constraint equations which exhibit the difference between
RCs and MCs admitted by the spacetime. Consider the following
plane symmetric static spacetime
\begin{equation}
ds^2=(ax+b)^2dt^2-dx^2-(cx+d)^2(dy^2+dz^2),
\end{equation}
where $a,b,c,d\in \Re,~ac\neq 0\neq ad-bc$. In this example, we
obtain five MCs in which three are the usual isometries and the
remaining two are proper MCs but the RCs are infinite dimensional.

The most general form of cylindrically symmetric static spacetime
is given by
\begin{equation}
ds^2=e^{\nu(r)}dt^2-dr^2-e^{\lambda(r)}d\theta^2-e^{\mu(r)}dz^2.
\end{equation}
The only non-zero components of the energy-momentum tensor turn
out to be $T_{00},~T_{11},~T_{22},~T_{33}$. We have found [3] that
the non-degenerate energy-momentum tensor gives either three,
four, five, six, seven or ten independent MCs in which three are
isometries and the rest are proper. There are four worth
mentioning cases where we have obtained the group of MCs
finite-dimensional even the energy-momentum tensor is degenerate,
i.e., either three, four, five or ten. It can be seen from the two
papers [3,13] that RCs and MCs become similar but the constraints
are different. Here we present examples by solving these
constraints which give different spacetimes for the two
collineations.

The following cylindrically symmetric metric
\begin{equation}
ds^2=\cosh^2crdt^2-dr^2-(\cosh cr)^{-1}d\theta^2-(\cosh
cr)^{-1}dz^2,
\end{equation}
where $c$ is an arbitrary constant, admits 4 MCs and also 4
isometries but it has 7 RCs. The spacetime
\begin{equation}
ds^2=(r/r_0)^{2a}dt^2-dr^2-(r/r_0)^{2a}d\theta^2-
(r/r_0)^{2a}dz^2,
\end{equation}
where $a$ and $r_0$ are arbitrary constants such that $a\neq 0,1$,
admits 10 MCs with 6 KVs but 7 RCs. Taking $\nu=\lambda=\mu$ in
Eq.(6), this metric admits 6 MCs and also 6 KVs but 7 RCs. The
following spacetime
\begin{equation}
ds^2=(\cosh cr)^{-1}dt^2-dr^2-\cosh^2crd\theta^2-(\cosh
cr)^{-1}dz^2.
\end{equation}
has 4 MCs and also 4 KVs but 7 RCs.

In this addendum, we have provided examples which clearly indicate
the difference of the symmetries for the Ricci and matter tensors.
Also, we have incorporated the missing references in the previous
papers [1-3]. It is mentioned here that RCs and MCs will exactly
be similar for those spacetimes where $R_{ab}=T_{ab}$ or
equivalently for vacuum spacetimes. For example, in the case of
Schwarzschild metric, every direction is RC/MC.

\vspace{2cm}

{\bf \large References}

\begin{description}

\item{[1]} Sharif, M.: J. Math. Phys. {\bf 44}(2003)5142.

\item{[2]} Sharif, M.: J. Math. Phys. {\bf 45}(2004)1518.

\item{[3]} Sharif, M.: J. Math. Phys. {\bf 45}(2004)1532.

\item{[4]} Sharif, M.: Nuovo Cimento {\bf B116}(2001)673.

\item{[5]} Sharif, M.: Astrophys. Space Sci. {\bf 278}(2001)447.

\item{[6]} Camc{\i}, U. and Sharif, M.: Gen Rel. and Grav. {\bf
35}(2003)97.

\item{[7]} Camc{\i}, U. and Sharif, M.: Class. Quant. Grav.
{\bf 20}(2003)2169-2179.

\item{[8]} Sharif, M. and Sehar Aziz: Gen Rel. and Grav. {\bf
35}(2003)1091.

\item{[9]} Tsamparlis, M. and Apostolopoulos, P.S.: Gen. Rel.
and Grav. {\bf 36}(2004)47.

\item{[10]} Ziad, M.: Gen. Relativ. Gravit. {\bf 35}(2003)915.

\item{[11]} Stephani, H., Kramer, D., MacCallum, M.A.H.,
Hoenselaers, C. and Hearlt, E.: {\it Exact Solutions of Einstein's
Field Equations} (Cambridge University Press, 2003).

\item{[12]} Farid, Taha Bin, Qadir Asghar and Ziad, M.: J. Math. Phys. {\bf
36}(191995)5812.

\item{[13]} Qadir, Asghar, Saifullah, K. and Ziad, M.:
Gen. Relativ. Gravit. {\bf 35}(2003)1927.

\end{description}

\end{document}